\documentstyle[aps,12pt,preprint,psfig]{revtex}

\begin{document}
\draft
\preprint{
\hbox{PKU-TP-98-54}}
\title{Large transverse momentum direct photon production in the coherent
diffractive processes at hadron colliders}
\author{Feng Yuan}
\address{{\small {\it Department of Physics, Peking University, Beijing 100871,
People's Republic of China}}}
\author{Kuang-Ta Chao}
\address{{\small {\it China Center of Advanced Science and Technology (World
Laboratory), Beijing 100080, People's Republic of China\\
and Department of Physics, Peking University, Beijing 100871, People's
Republic of China}}}
\maketitle

\begin{abstract}
Direct photon production at large transverse momentum in the coherent
diffractive processes at hadron colliders is calculated in the two-gluon
exchange model. We find that the amplitude for the production process is
related to the differential off-diagonal gluon distribution function in the
proton. We estimate the production rate at the Fermilab Tevatron by
approximately using the usual gluon distribution function. Because of the
clean signature, this process can be used to detailed study the small-$x$
physics, and the coherent diffractive processes at hadron colliders.
\end{abstract}

\pacs{PACS number(s): 12.40.Nn, 13.85.Ni, 14.40.Gx}

\preprint{\ \hbox{PKU-TP-98-54}}

In recent years, there has been a renaissance of interest in diffractive
scattering. These diffractive processes are described by the Regge theory in
terms of the Pomeron ($I\!\!P$) exchange\cite{pomeron}. The Pomeron carries
quantum numbers of the vacuum, so it is a colorless entity in QCD language,
which may lead to the ``rapidity gap" events in experiments. However, the
nature of the Pomeron and its interaction with hadrons remain a mystery. For
a long time it had been understood that the dynamics of the ``soft Pomeron''
was deeply tied to confinement. However, it has been realized now that much
can be learned about QCD from the wide variety of small-$x$ and hard
diffractive processes, which are now under study experimentally.

On the other hand, as we know that there exist nonfactorization effects in
the hard diffractive processes at hadron colliders \cite
{preqcd,collins,soper,tev}. First, there is the so-called spectator effect
\cite{soper}, which can change the probability of the diffractive hadron
emerging from collisions intact. Practically, a suppression factor (or
survive factor) ``$S_F$'' is used to describe this effect. Obviously, this
suppression factor can not be calculated in perturbative QCD, which is now
viewed as a nonperturbative parameter. Typically, the suppression factor
$S_F $ is determined to be about $0.1$ at the energy scale of the Fermilab
Tevatron\cite{tev}. Another nonfactorization effect discussed in literature
is associated with the coherent diffractive processes at hadron colliders
\cite{collins}, in which the whole Pomeron is induced in the hard
scattering. It is proved in \cite{collins} that the existence of the leading
twist coherent diffractive processes are associated with a breakdown of the
QCD factorization theorem.

In this paper, we will calculate the direct photon production at large
transverse momentum in the coherent diffractive processes at hadron
colliders under the framework of the two-gluon exchange parametrization of
the Pomeron. As shown in Fig.1, the whole Pomeron represented by the
color-singlet two-gluon system emits from one hadron and interacts with
another hadron to produce photon plus a quark jet. The large transverse
momentum is required to ensure the validity of the application of
perturbative QCD. In the leading order of QCD, there is only the quark
initiated partonic process contributing to the coherent diffractive
production of large transverse momentum photon. The partonic process
$qp\rightarrow q\gamma p$ is shown in Fig.2.

To calculate the cross section for the partonic diffractive process, we use
the two-gluon exchange parametrization of the Pomeron model. The two-gluon
exchange model has been used to calculate diffractive photoprodution
processes including the productions of vector meson\cite{th1}, open charm
\cite{th2}, large $p_T$ di-jet\cite{th3}. This model has gained some success
in the description of these processes at $ep$ colliders\cite{hera-ex}.
Recently, we have extended this model to calculate the diffractive processes
at hadron colliders. We have calculated the diffractive $J/\psi $\cite{psi}
production, charm jet \cite{charm} production, and massive muon pair and $W$
boson productions\cite{dy} at hadron colliders. An important feature of this
two-gluon exchange model is the relation of the production amplitude to the
off-diagonal gluon distribution function \cite{offd}. The results of \cite
{psi,charm,dy} indicate that we can explore much low $x$ gluon distribution
in the proton by studying the diffractive processes in hadron collisions.

In the two-gluon exchange model, the leading order contribution to the
partonic diffractive process $qp\rightarrow \gamma qp$ comes from the four
diagrams shown in Fig.2. These four diagrams are the same as those
calculated in \cite{dy} except the difference on the virtuality of the
photon. In Ref.\cite{dy} the produced photon is virtual and time-like, while
here in this paper we will calculate the production of real photon in the
diffractive process at large transverse momentum. Also, these four diagrams
imply that the partonic process $qp\rightarrow \gamma qp$ is related by
crossing to diffractive di-quark jets photoprodution process $\gamma
p\rightarrow q\bar qp$\cite{th3}.

As indicated in Refs.\cite{th3,light}, in the calculations of the amplitude
for the massless particle production in the diffractive processes by using
the two-gluon exchange model there is no contribution from the region of
$l_T^2<k_T^2$, where $l_T^{}$ and $k_T^{}$ are the transverse momenta of the
loop momentum and the final state photon momentum as shown in the diagrams
of Fig.2. This is contrast to the results of Refs.\cite{psi,charm,dy}, where
the dominant (large logarithmic) contribution comes from the small $l_T^2$
region. Therefore, the expansion method used in of \cite{dy} in which $l_T^2$
is viewed as a small parameter is not further applicable for the calculation
of the real photon production in the diffractive process $qp\rightarrow
\gamma qp$. So, in the following we directly calculate the cross section by
squaring the amplitude and picking up the dominant terms.

Due to the positive signature of these diagrams (color-singlet exchange), we
know that the real part of the amplitude cancels out in the leading
logarithmic approximation. To evaluate the imaginary part of the amplitude,
we must calculate the discontinuity represented by the crosses in each
diagram of Fig.2.

In our calculations, we express the formula in terms of the Sudakov
variables. We select $q$ and $p$ as the light cone momenta, where $q$ and $p$
are the momenta of the incident quark and the diffractive proton
respectively. For high energy hadron scattering, we know that the light
quark and the proton masses are much smaller than the hard scattering scale.
So, we have $q^2=0,~p^2=0$, and we set $2pq=s,$ where $s$ is the total c.m.
energy of the quark-proton system, i.e, the invariant mass of the partonic
process $qp\rightarrow \gamma qp$. Thus, every four-momenta $ki$ can be
decomposed as 
\begin{equation}
k_i=\alpha _iq+\beta _ip+k_{iT},
\end{equation}
Where $\alpha _i$ and $\beta _i$ are the momenta fractions of $q$ and $p$
carried by $k$. $k_{iT}$ is the transverse momentum, and it satisfies 
\begin{equation}
k_{iT}\cdot q=0,~~~k_{iT}\cdot p=0.
\end{equation}

All of the Sudakov variables for every momentum are determined by the
on-shell conditions of the external lines and the crossed lines in each
diagram. For example, the Sudakov variables associated with the momenta $k$
and $u$ are, 
\begin{eqnarray}
\alpha _u &=&0,~~\beta _u=x_{I\!P}=\frac{M_x^2}s,~~u_T^2=0 \\
\alpha _k(1+\alpha _k) &=&-\frac{k_T^2}{M_X^2},~~\beta _k=-\alpha _k\beta _u,
\end{eqnarray}
where $k_T$ is the transverse momentum of the outgoing photon, $M_x^2$ is
the invariant mass of the diffractive final state (including the large
transverse momentum photon and the quark jet).

For the high energy diffractive process, we know that $M_x^2\ll s$, i.e., we
have $\beta _u$ ($x_{I\!P}$) as a small parameter. In the following
calculations of the cross section for the partonic process, we take the
leading order contribution and neglect the high order contributions which
are proportional to $\beta _u=\frac{M_x^2}s$.

By using the above Sudakov variables, we can evaluate the diffractive cross
section formula for the partonic process $qp\to \gamma qp$ as 
\begin{equation}
\frac{d\hat \sigma (qp\rightarrow \gamma qp)}{dt}|_{t=0}=\frac{
dM_X^2d^2k_Td\alpha _k}{16\pi s^216\pi ^3M_X^2}\delta (\alpha _k(1+\alpha
_k)+\frac{k_T^2}{M_X^2})\sum \overline{|{\cal A}|}^2,  \label{xs}
\end{equation}
where ${\cal A}$ is the amplitude of the process. We know that the real part
of the amplitude does not contribute. And the imaginary part of the
amplitude ${\cal A}$ has the following form for every diagram of Fig.2, 
\begin{equation}
{\rm Im}{\cal A}=C_F\int \frac{d^2l_T}{(l_T^2)^2}F\times \bar u_i(u-k)\Gamma
_\mu v_i(q),  \label{ima}
\end{equation}
where $C_F=\frac 29$ is the color factor for the four diagrams. $F$ is
defined as 
\begin{equation}
F=\frac 3{2s}g_s^2ee_qf(x^{\prime },x^{\prime \prime };l_T^2),
\end{equation}
where 
\begin{equation}
f(x^{\prime },x^{\prime \prime };l_T^2)=\frac{\partial G(x^{\prime
},x^{\prime \prime };l_T^2)}{\partial {\rm ln}l_T^2}.  \label{offd1}
\end{equation}
The function $G(x^{\prime },x^{\prime \prime };l_T^2)$ is the off-diagonal
gluon distribution function in the proton\cite{offd}, where $x^{\prime }$
and $x^{\prime \prime }$ are the momentum fractions of the proton carried by
the outgoing and returning gluons of each diagrams of Fig.2.

After a straightforward calculation, and neglecting the higher order
contributions which are proportional to $\beta _u=\frac{M_x^2}s$, we get the
amplitude squared as 
\begin{eqnarray}
\sum \overline{|{\cal A}|^2} &=&\frac{128\pi ^5\alpha _s^2\alpha e_q^2}9s^2
\frac{(1+\alpha ^2)k_T^2}{\alpha _k(M_X^2)^2}[\frac 1{\pi ^2}\int \frac{
d^2l_T}{(l_T^2)^2}\frac{d^2l_T^{\prime }}{(l_T^{\prime 2})^2}f(x^{\prime
},x^{\prime \prime };l_T^2)f(y^{\prime },y^{\prime \prime };l_T^{\prime 2}) 
\nonumber \\
&&\frac{(1+\alpha _k)^2l_T^2-(1+\alpha _k)(k_T,l_T)}{(\vec k_T-(1+\alpha _k)
\vec l_T)^2}\frac{(1+\alpha _k)^2l_T^{\prime 2}-(1+\alpha
_k)(k_T,l_T^{\prime })}{(\vec k_T-(1+\alpha _k)\vec l_T^{\prime })^2}],
\end{eqnarray}
where $f(x^{\prime },x^{\prime \prime };l_T^2)$ and $f(y^{\prime },y^{\prime
\prime };l_T^{\prime 2})$ are the differential off-diagonal gluon
distribution functions associated with the loop momenta $l_T^{}$ and
$l_T^{\prime }$ respectively.

From the above equations, we can see that the amplitude for the partonic
process $qp\to \gamma qp$ is related to the off diagonal gluon distribution
in the proton. By now there is no parametrization of the off-diagonal parton
distributions, and it is expected that at small $x$ the off-diagonal gluon
distribution is not far away from the usual diagonal gluon distribution\cite
{off-diag}. So in the following, we approximate the off-diagonal gluon
distribution by the usual gluon distribution, i.e., $G(x^{\prime },x^{\prime
\prime };Q^2)=G(x;Q^2)$ and $f(x^{\prime },x^{\prime \prime };Q^2)=f_g(x;Q^2)
$, where $x=x_{I\!P}=\frac{M_X^2}s$ . $f_g(x;Q^2)$ is the usual differential
gluon distribution in the proton.

So, after integrating over the azimuth angle of $\vec l_T$, we get 
\begin{equation}
\sum \overline{|{\cal A}|^2}=\frac{128\pi ^5\alpha _s^2\alpha e_q^2}9s^2
\frac{(1+\alpha ^2)k_T^2}{\alpha _k(M_X^2)^2}|{\cal I}|^2,
\end{equation}
where the integration ${\cal I}$ is defined as 
\begin{equation}
{\cal I}=\int \frac{dl_T^2}{(l_T^2)^2}\frac 12[1-\frac{k_T^2-(1+\alpha
_k)^2l_T^2}{|k_T^2-(1+\alpha _k)^2l_T^2|}]f_g(x;l_T^2).
\end{equation}
From the above results of the integration, we can see that the amplitude
${\cal A}(qp\rightarrow q\gamma p)$ will be zero in the integral region of
$l_T^2<k_T^2/(1+\alpha _k)^2$. So, the dominant contribution to the
integration of the amplitude over $l_T^2$ will come from the region of
$l_T^2\sim k_T^2/(1+\alpha _k)^2$. This behavior of the amplitude integration
over $l_T^2$ is similar to the results of Refs.\cite{th3,light}. Therefore,
if we neglect the evolution effects of the gluon distribution function in
this region, the integration will then be 
\begin{equation}
{\cal I}=\frac{(1+\alpha _k)^2}{k_T^2}f_g(x;\frac{k_T^2}{(1+\alpha _k)^2}).
\end{equation}
Under this approximation, the differential cross section for the partonic
process $qp\rightarrow \gamma qp$will then be 
\begin{eqnarray}
\frac{d\hat \sigma }{dt}(qp\rightarrow \gamma qp)|_{t=0}=\int 
&&_{M_X^2>4k_T^2}dM_X^2dk_T^2d\alpha _k\frac{\pi ^2\alpha \alpha _s^2e_q^2}{
18}\delta [\alpha _k(1+\alpha _k)+\frac{k_T^2}{M_X^2}]  \nonumber \\
&&\ \ \frac{(1+\alpha _K^2)(1+\alpha _K)^2}{\alpha _k(M_X^2)^2k_T^2}(f_g(x;
\frac{k_T^2}{(1+\alpha _k)^2}))^2.  \label{dxs}
\end{eqnarray}

From Eq.(\ref{dxs}), we can see that the cross section for the partonic
process $qp\rightarrow \gamma qp$ is proportional to the square of the
differential gluon distribution function in the proton, which is similar to
the results of \cite{th3,light}. However, in the processes of Refs.\cite
{th1,th2,psi,charm,dy}, the cross section is proportional to the integrated
gluon distribution function. This is because in the processes of \cite
{th1,th2,psi,charm,dy}, there exists large logarithmic contribution in the
region of $l_T^2\ll M_x^2$, which will lead to the gluon distribution after
integrating over $l_T^2$ in the small $l_T^2$ region. On the other hand, in
the processes of \cite{th3,light} and the direct photon production process
calculated in this paper, in the smaller $l_T^2$ region, ($l_T^2<\frac{k_T^2
}{(1+\alpha _u^2)}$), the integral is equal to zero, so there is no large
logarithmic contribution to the amplitude. The contribution from the
integration of large $l_T^2$ region only leads to the differential gluon
distribution terms, because the integral decreases rapidly as $l_T^2$
increases.

Using Eq.~(\ref{dxs}), we can estimate the production rate of large $p_T^{}$
direct photon in the diffractive processes at the Fermilab Tevatron. We
adopt the value of $S_F=0.1$ to describe the spectator effect at this
machine. The numerical results are displayed in Fig.3 and Fig.4. In the
calculations, we set the scale of the running coupling constant identical to
the scale of the gluon distribution function, i.e, $Q^2=\frac{k_T^2}{
(1+\alpha _K^2)}$. For the parton distribution functions, we select the GRV\
NLO set\cite{grv}.

In Fig.3, we plot the differential cross section as a function of the lower
cut of the transverse momentum of the produced photon, $k_{T\min }^{}$.
Approximately, we can derive the dependent on $k_{T\min }$ from Eq.(\ref{dxs}
), in which the integration over $M_X^2$ mainly comes from the region of
$M_X^2\sim 4k_T^2$. So the differential cross section behaves as,
\[
\frac{d^2\hat \sigma }{dk_T^2dt}|_{t=0}\sim \frac 1{(k_T^2)^3}
(f_g(x;4k_T^2))^2 
\]
Where $x=4k_T^2/M_X^2$. Similarly, integration over $k_T^2$ dominantly comes
from the region of $k_t^2\sim k_{t\min }^2$. So we get the approximate
dependence of the cross section on $k_{T\min }$ as 
\[
\frac{d\hat \sigma }{dt}|_{t=0}\sim \frac 1{(k_{T{\rm \min }}^2)^2}
(f_g(x;4k_{T{\rm \min }}^2))^2 
\]
This dependence can be seen from Fig,3.\ This is a distinctive feature of
the calculation of our model for this process.

In Fig.4, we plot the dependence of the differential cross section on
$x_{1\min }$, $x_1$ is the longitudinal momentum fraction of the proton
carried by the incident quark. $x_{1\min }$ is the lower bounder of $x_1$ in
the integration of the cross section. From this figure we can see that the
dominant contribution to the cross section comes from the region of $x_1\sim
10^{-1}$, which is similar to the case of the diffractive massive muon pair
and $W$ boson productions \cite{dy}. However, if we compare this result to
that of the diffractive $J/\psi $ and charm jet productions \cite{psi,charm}
, we will find that the dominant contribution region of $x_1$ to the cross
section of direct photon production here is some orders of magnitude larger
than that of $J/\psi $ and charm jet productions. This is because the direct
photon production is the quark initiated process which is sensitive to
large-$x$ quark distribution in the proton. However, the productions of $J/\psi $
and charm jet are the gluon initiated processes, so they are sensitive to
the small-$x$ gluon distribution in the proton. So, the $x_1$ dependence of
these two types of processes are distinctive different.

In conclusion,we have shown that the large transverse momentum direct photon
diffractive production can provide a test for the two-gluon exchange model
of the Pomeron, and can also be viewed as a compensate to the $J/\psi $ and
charm jet productions for the study of the coherent diffractive processes at
hadron colliders. And furthermore, because of the clean signature, this
process can be used to detailed study the small-$x$ physics, and the
coherent diffractive processes at hadron colliders. The photon production
process is a quark induced process, and is sensitive to the large-$x$ quark
distribution function in the proton. However, all of the coherent
diffractive processes calculated in the two-gluon exchange model are related
to the off-diagonal gluon distribution function in the proton. So, the
measurement of these processes at hadron colliders may provide a lot of
information about the off-diagonal gluon distribution function.

This work was supported in part by the National Natural Science Foundation
of China, the State Education Commission of China, and the State Commission
of Science and Technology of China.

\newpage
\vskip 10mm \centerline{\bf \large Figure Captions} \vskip 1cm \noindent
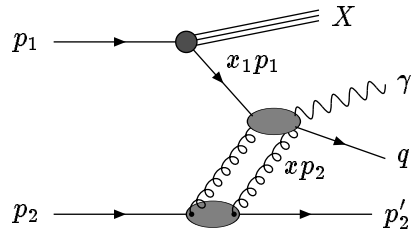
Fig.1. Sketch diagram for the diffractive direct photon production at hadron
colliders in perturbative QCD.

\noindent
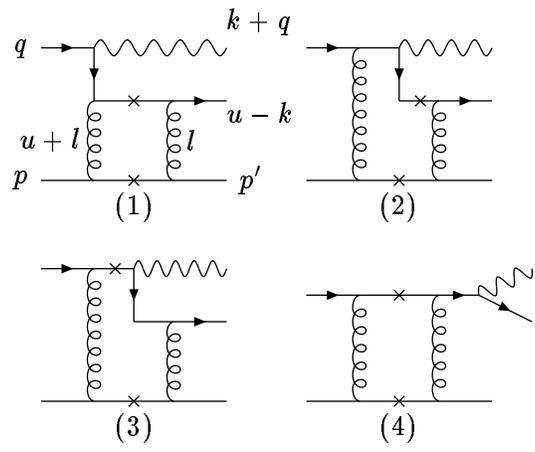
Fig.2. The lowest order perturbative QCD diagrams for partonic process
$qp\rightarrow \gamma qp$.

\noindent
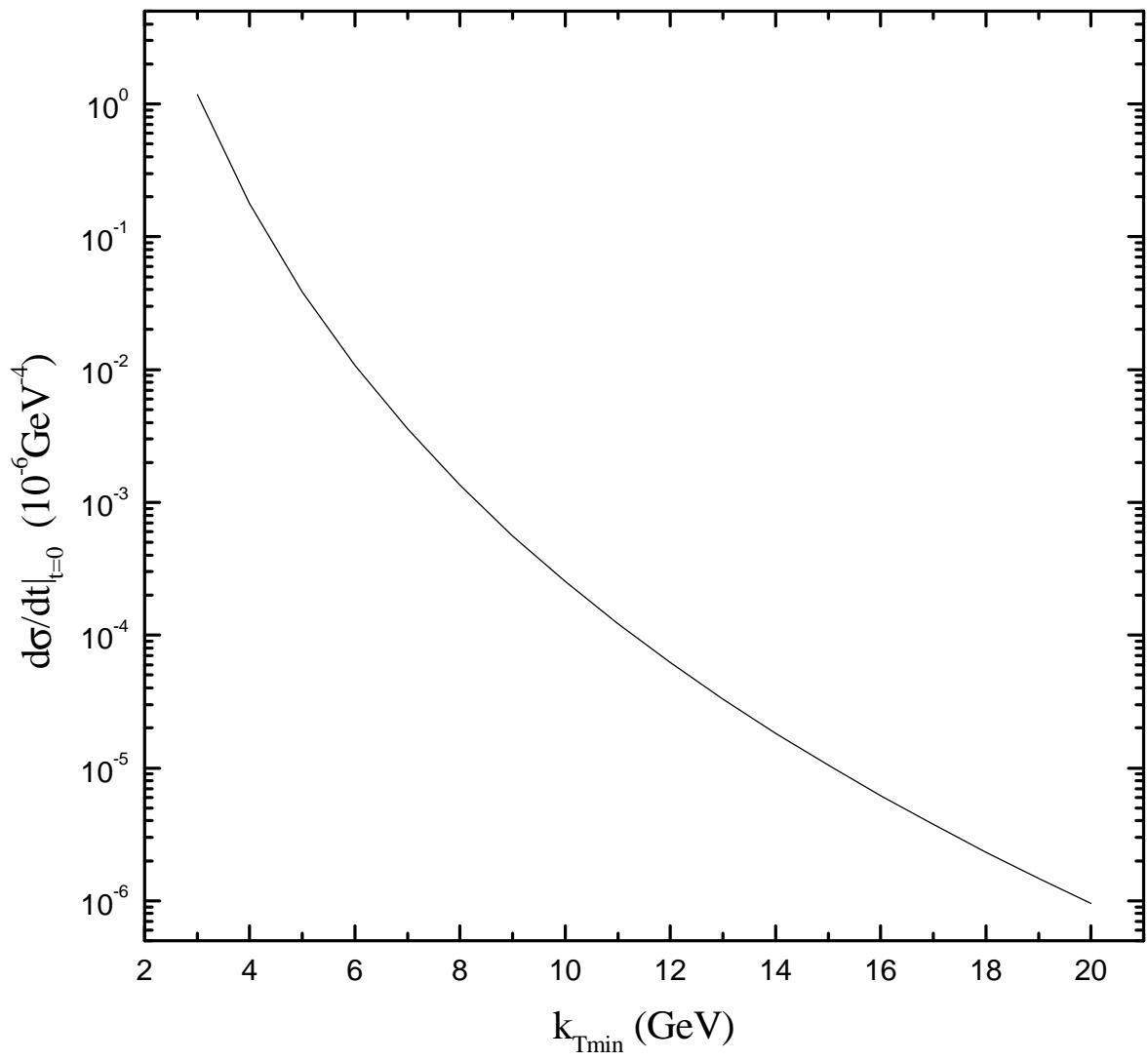
Fig.3. The differential cross section $d\sigma /dt|_{t=0}$ for the large
transverse momentum direct photon production at the Fermilab Tevatron as a
function of $k_{T{\rm min}}$, where $k_{T{\rm min}}$ is the lower bound of
the transverse momentum of the produced photon.

\noindent
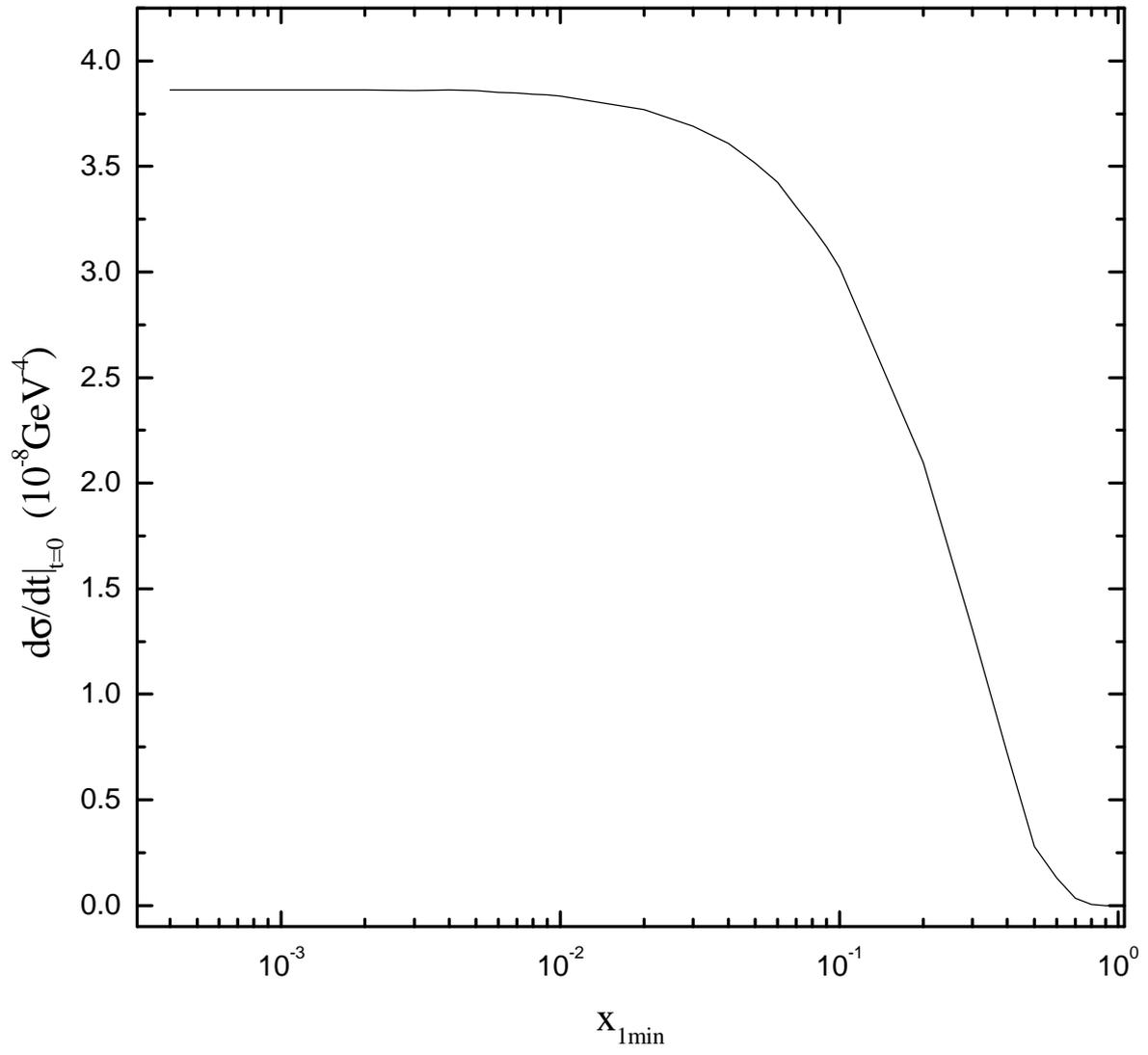
Fig.4. The differential cross section $d\sigma /dt|_{t=0}$ as a function of
$x_{1{\rm min}}$, where $x_{1{\rm min}}$ is the lower bounder of $x_1$ in the
integration of the cross section.

\end{document}